# SOLARNET Metadata Recommendations for Simulated Data

*Version 2.2-live – 18. March 2023*

*Stein Vidar Hagfors Haugan and Terje Fredvik*

This document has received funding from the European Union's Horizon 2020 and FP7 programmes under grant agreements No 824135 and 31295. Version 2.0 of this document was the final version produced under these grants, but it will keep evolving as further needs and issues arise. The *latest working copy* of the document can be found at https://sdc.uio.no/open/solarnet/. Please use that version when adding comments/suggested changes (using track changes), before sending to prits-group@astro.uio.no

## *Preface*

Until the advent of the SOLARNET recommendations, metadata sharing of simulated data within the Solar Physics community has been mostly on a "private communication" basis, with the description of the data format and content conveyed in an *ad hoc* manner. This document aims to amend this situation by establishing recommendations for representing such data and the associated metadata.

Within observational Solar physics, the most common file format is Flexible Image Transport Format (FITS – see FITS Standard 4.0). It is also the file format used to share BIFROST simulations and derived products (e.g. synthesised observables) through the Hinode archive at ITA/UiO. Additionally, most visualization software within Solar physics use FITS files as input. Therefore, these metadata recommendations are expressed here through FITS keywords. *However, this does not preclude the use of other file formats expressing the same metadata in different forms*.

In this document, "simulated observations" refers to any "observation-like" data product derived from the simulations themselves, defined over a two-dimensional footprint of a simulation. Typically, simulated observations are normal observables such as synthetic spectra or images, but they may also represent e.g. the height at which $\tau$=1. Any three-dimensional derived products should normally be treated on par with actual simulation data.

In order to facilitate reuse of analysis and visualization software written for true Solar observations, simulated observations should mimic true observations to a high degree.

For that reason, *almost all of the SOLARNET Metadata Recommendations ([S-META] hereafter applies to simulated observations*, thus for these we will here only mention some of the applicable deviations and additions, give some clarifying remarks, and reiterate a few important points. The notation used is the same as in [S-META] (e.g. FITS keyword names and values are in bold monospace font).

# Table of Contents

# Part A.  Simulation data

## 1. Introduction

Solar simulation data are fundamentally different from observational data – they typically have 3 spatial dimensions rather than 2, and represent a wide range of physical parameters that are not observables.

Also, simulation data volumes are often orders of magnitude larger than both real and simulated observations, and the data are often defined on a complex, non-rectangular and non-uniform grid (staggered mesh, varying spatial resolution with height, etc) that is specific to each simulation code.

Thus, to facilitate use of the data by other groups it will often be necessary to publish downsampled and non-irregular versions of the simulation data rather than the raw simulation data. At the moment, we recommend using the FITS file format for publishing such data. For general information in this regard, see [S-META].

Note that "non-irregular" does not mean a constant resolution throughout the simulation – coordinates may be tabulated to specify increasing or decreasing resolution in one or more dimensions.

## 2. General considerations

HDUs containing simulation data (Sim-HDUs) must have `SOLARNET=0.5` or `1.` and may use the keyword `SOLNETEX` to exclude certain keywords from being interpreted within the `SOLARNET` metadata framework. It should have `SIM_HDU=1` (not `OBS_HDU`). All Sim-HDUs must have a unique `EXTNAME`.

Any HDU with a nonzero value of `SOLARNET` may use the mechanisms described in the Appendices of [S-META] (e.g., variable keywords, pixel lists and meta-HDUs).

## 3. WCS and related keywords for simulation data

### 3.1. Spatial coordinates

Simulations normally have no specific location, nor an observer position. Furthermore, a determination of a simulation's height relative to the Solar surface in absolute terms may be impossible or near meaningless. Therefore, using standard WCS coordinates would be an unnecessary distraction from the most important aspects of the data. Nevertheless, the WCS system provides a standardized framework for numeric specifications of any set of coordinates for the data cube contained in an HDU.

For simulation data we recommend using a simple, right-handed Cartesian coordinate system labelled through `CTYPEi` as `(x, y, z)`, where `z` is the height, all expressed in (mega-)meters. If the physical orientation of the x and y axes have any meaning at all relative to the Sun, the y coordinate axis should point towards Solar North. If the data cube dimensions are oriented differently, the `PCi_j` transformation matrix can easily be used to align the coordinates correctly.

However, some simulations relate directly to an actual observation or feature. In such cases, the spatial relationship between the simulation and the observation/feature should be defined by setting `HGLN_OBS` (observer's Heliographic longitude) and `HGLT_OBS` (observer's Heliographic latitude), to match the Heliographic longitude and latitude of the pixel with coordinates `(x,y,z) = (0,0,0)`. Both `HGLN_OBS` and `HGLT_OBS` default to zero, placing the simulation on the equator. Then, if possible, conventional WCS coordinates for the data cube may be given in an alternate coordinate system, with the `a` in `CRPIXja`, `CDELTia`, and `CRVALia` set to some letter `A-Z`.

For large-footprint simulations where a spherical coordinate system has been used internally, it should be fairly easy to use the standard WCS coordinates `HGLN/HGLT` (heliographic longitude and latitude) plus `HECR` (radial distance from sun centre) or `HECH` (radial distance from the solar surface). `RSUN_REF` is used to give the solar radius used in coordinate specifications (thus providing ample room for varying definitions), and a sufficiently large value for `CSYERi` (systematic error in coordinate i) can be used to reflect the (possibly large) uncertainty in the radial coordinate.

## 3.2. Tabulation of coordinates with varying resolution

Often, simulation data have varying resolution with height, necessitating a tabulation of the `z` coordinate. This can be achieved in a standardized way using the `-TAB` coordinate tabulation mechanism described in Paper III, Section 6.1.3

As a simple (but normally sufficient) example, we assume that a data cube with dimensions `(x,y,z) = (200,200,300)` where the z coordinate requires tabulation. The first two coordinates are specified as usual, but the third, tabulated z coordinate would be specified with:

```
CTYPE3   = 'z----TAB'      / Coord. 3 is z, tabulated
PS3_0    = 'z-coordinate'  / EXTNAME of binary table ext. containing z-coordinate
PS3_1    = 'values'        / TTYPEn value of bintable cell containing z-coordinate
```

Note the four dashes – this is to make the `CTYPE3` value adhere to the '4-3' pattern, which may be necessary for some software to recognize the tabulated nature of the coordinate.

The binary table extension with `EXTNAME='z-coordinate'` should contain a column named 'values', containing a 300-element array (i.e., `TFORM1='(300)'`).

Here we have not specified `PV3_1` (`EXTVER` of the binary table containing the z coordinate values) nor `PV3_2` (`EXTLEVEL` of the binary table), all of which default to 1. We have also not specified `CRPIX3`, `CDELT3` nor `CRVAL3`, as they default to 0, 0, and 1 respectively, meaning that a lookup of `z` for pixel `(*,*,k)` will be done at index `k` in the single-cell array with `TTYPEn='values'`, i.e. a one-to-one correspondence.

## 4. Time-related keywords

Simulation data are typically stored with one snapshot per file, but snapshots should have an accurate time specification that is consistent between them, indicating the passage of Solar time in the simulation. This should be done by setting `DATE-BEG` in the first snapshot (somewhat arbitrarily) to the wall time at the start of the simulation, with `DATE-BEG` in later snapshots reflecting the elapsed (Solar) time since the start.

However, if a simulation attempts to model the physics underlying *an actual* observation or event, the `DATEREF` keywords should of course be set to line up the simulation's time coordinate with that of the observation/event.

If a Sim-HDU contains a time dimension/time coordinate (`CTYPEi='UTC'`), `DATEREF` (the time at which the `UTC` coordinate is zero), `CRPIXj`, `PCi_j` (if applicable) `CDELTi` and `CRVALi` must be specified such that the time coordinate reflects the elapsed Solar time since the beginning of the observation.

For each snapshot, `ELAPSED` should be set equal to the number of seconds in Solar time elapsed since the beginning of the simulation. `SNAPSHOT` should be set to the snapshot number.

To tie snapshots in individual files together in a more formal way, the Meta-HDU mechanism described in [S-META] can be used. This mechanism may also be used in order to split simulations along other dimensions if that seems necessary to have reasonable file sizes.

## 5. Description of the data contents

### 5.1. Data type/units (BTYPE/BNAME/BUNIT)

As for regular observations, the keywords `BTYPE`, `BNAME`, and `BUNIT` should be used to describe the nature of the data, see [S-META] Section 15.4 "Mandatory data description keywords" and [S-META] Section 18.5 "Optional data description keywords". Note, however, that the recommendations given in this section and Section 15.1, differ slightly from the recommendations given in [S-META].

The notation of mathematical expressions in `BUNIT` and `BNAME` should follow the rules in Table 6 of the FITS Standard, e.g., "`log(x)`" is defined as the common logarithm of `x` (to base 10).

Note that any mathematical operations such as logarithms must be included in `BUNIT`, e.g., `BUNIT='log(kg m^(-3))'`.

`BTYPE` will typically be used as an axis label when plotting the value as a function of spatial position or time and will often also be used as part of the file name containing the data.

`BNAME` should be used to give a "human readable" explanation of the contents. Based mostly on convention from the published Bifrost FITS files, we recommend the values below for various types of data:

```
BTYPE           BNAME
lgrho           log(mass density)
ux              bulk velocity in x
uy              bulk velocity in y
uz              bulk velocity in z
bx              magnetic field strength in x
by              magnetic field strength in y
bz              magnetic field strength in z
lgpg            log(gas pressure)
lgtg            log(gas temperature)
lgne            log(electron density)
lgH1            log(population density in ground state of hydrogen)
lgH2            log(population density in n=2 state of hydrogen)
lgH3            log(population density in n=3 state of hydrogen)
:
lgHx            log(population density in n=x state of hydrogen)
lgnp            log(proton density)
lgeta_amb       log(ambipolar diffusion eta_amb)
lgeta_art       log(artificial diffusion eta_art)
lgq_jamb        log(joule heating from ambipolar diffusion)
```

## 5.2. Binning/sampling factors

Although the FITS data cubes with simulation data typically are not representative of the internal simulation data format, it is in any case useful to have some information regarding the relative resolution of the data and the simulations. Thus, the relative resolution of the internal format and the FITS data cube in each dimension should be reported in NBINj, and the total binning factor should be reported in NBIN. If sampling has been performed instead, use NSAMPj and NSAMP. in the case of non-uniform binning/sampling, the average binning/sampling factors should be used.

## 5.3. Cadence

The average time step between snapshots should be reported in CADENCE.

## 5.4. Instrument/data characteristics etc. [not applicable]

## 5.5. [Placeholder] Quality aspects [not applicable]

## 5.6. Data statistics and other properties

Of the keywords defined in [S-META] Section 5.6, only DATAMEAN makes sense for most simulations, but the others may also be used. It would normally make sense to use the variable-keyword mechanism: E.g., with a data cube with dimensions [x,y,z]=[Nx,Ny,Nz], having a DATAMEAN variable keyword with dimension [1,1,Nz] specifies that the average is taken over each (x,y) plane.

The $\tau = 1$ height at 500nm should be specified with ZTAU51, if available

Of course, a single snapshot does not have an actual cadence, but the keyword `CADENCE` should be set to the "local" cadence to indicate the time between snapshots, perhaps the average of the time difference to the previous and subsequent snapshots. `SNAPSHOT` should be set to the snapshot number. `ELAPSED` should be used to indicate the number of seconds of simulated time has passed since the start of the simulation.

## 6. Metadata about affiliation, origin, acquisition, etc.

Note: Here we reuse a number of keywords defined in [S-META] to keep the number of keywords down between the two documents, even though some may be somewhat of a misnomer for simulations. This is a "tweaked" list of those mentioned there that may apply to simulations:

- `PROJECT`: Name(s) of the project(s) affiliated with the data
- `OBSRVTRY`: Name of the institution
- `CREATOR`: Name of the simulation code (to specify further details, see Sect. 8.1 in [S-META])
- `INSTRUME`: Alias for `CREATOR`
- `CAMERA`: Name of the software used to convert data from the internal format into FITS format, for further details use `PRxxxxn` keywords (see Sect. 8 in [S-META]) describing the conversion as a separate processing step.
- `SETTINGS`: Input parameters can be given as '`parameter1=n, parameter2=m`' (example below)
- `OBSERVER`: Who ran the code.
- `CAMPAIGN`: Use when the simulation is part of a specific effort/subproject
- `TARGET`: A comma-separated list of the targets of the simulation. E.g., '`Quiet Sun`', '`Filament`', '`Active Region`', '`Sunspot`', '`Umbra`', '`Penumbra`', '`Coronal Hole`', '`CME`', '`Flare`', etc.
- `SMLATION`: Unique identifier of the simulation, e.g., "run name"

We suggest that some effort is made to make connections between observational metadata and simulation metadata, reusing keywords used for observational data when a reasonable connection or analogy is possible. This will make it somewhat easier to reuse SVO designs for simulation archives.

Since the number of input parameters reported in `SETTINGS` may be quite high, we point out that the `CONTINUE` long string convention may be used to make the FITS header more readable, as in the example below. Strings should be enclosed with double quotes, and arrays may be specified using bracket notation:

```
SETTINGS = 'SOME_PARAMETER=5,        &' / Comment for SOME_PARAMETER
CONTINUE = 'ANOTHER_SETTING=8.5,     &' / Comment for ANOTHER_PARAMETER
CONTINUE = 'THIRD_INPUT="ALPHA",     &' / Comment for THIRD_INPUT_VALUE
CONTINUE = 'FOURTH=[5.2, 3.7, 8.6]   &' / Comment for FOURTH
CONTINUE = 'LAST=0  '                    / Comment for LAST_INPUT_VALUE
```

## 7. Grouping

For one-off simulations, `POINT_ID` can be identical to the `SMLATION` keyword, but for simulations that are part of an overarching series of simulations, `POINT_ID` should be set to an identifier of that overarching series, so it can be separated from other series.

## 8. Pipeline processing applied to the data

See Section 8 of [S-META], *including the footnotes*.

## 9. Integrity and administrative information

See Section 9 of [S-META].

## 10. Reporting of events

See corresponding section in [S-META] and note that features and events detected inside the simulation can be pinpointed using the pixel list mechanism.

# Part B. Simulated observations

## 11. Background and file formats

Simulated observations are in many respects similar to observational data, but in other respects they are similar to actual simulations (at least w.r.t metadata). The data content is in most cases identical to real observations (intensity as a function of two spatial dimensions, wavelength, and time) but may also represent non-observable quantities such as the physical height of $\tau_{500}$.

To facilitate analysis of simulated observations by the solar observations community using standard software, it is important that simulated observations masquerade as true observations to a sufficient degree. Thus, arguments regarding file formats for observations apply also to simulated observations: FITS is the current recommendation, but the metadata descriptions here apply to any format that might be applicable in the future.

For readability, we will use the term "sim-observation(s)" to mean "simulated observation(s)".

## 12. Header and Data Units (HDUs) in FITS files

For a background regarding header and data units (HDUs), see [S-META] Section 2.

As for real observations, HDUs containing sim-observations must have `SOLARNET=0.5` or `1`, `OBS_HDU=1`, and may use the keyword `SOLNETEX` to exclude certain keywords from being interpreted within the `SOLARNET` metadata framework. All Obs-HDUs must have a unique `EXTNAME`.

Any HDU with a nonzero value of `SOLARNET` may use the mechanisms described in the Appendices of [S-META] (e.g., variable keywords, pixel lists and meta-HDUs).

## 13. The World Coordinate System (WCS) and related keywords

Given the recommendations in Section 3.1, the easiest way to specify spatial coordinates for sim-observations is to carry over the coordinates `CTYPEi = (x, y)` from the simulation. This implicitly defines the observer position to be at an infinite distance directly above `(x, y) = (0, 0)`.

However, it is also possible to specify an explicit observer position with `HGLN_OBS`/`HGLT_OBS`/`DSUN_OBS` and use regular WCS coordinates such as `HPLN`/`HPLT` (Helioprojective longitude and latitude) as spatial coordinates. Such positioning makes it possible to represent sim-observations that have been made from another vantage point than straight above, since simulations are placed either implicitly at `(HGLN,HGLT)=(0,0)` or explicitly somewhere else (when `HGLN_OBS`/`HGLT_OBS` are specified in the simulation).

## 14. Time-related keywords

Time-related keywords such as `DATE-BEG`, `DATEREF`, `CRPIXj`, `PCi_j`, `CDELTi`, `CRVALi` (for `CTYPEi='UTC'`, if any), `ELAPSED` and `SNAPSHOT` should of course be set based on the values in the simulation. See also Section 4, and Section 4 of [S-META] if the data has a time coordinate.

## 15. Description of data contents

Much of the corresponding Section 5 in [S-META] does not apply to most sim-observations, except when mimicking the characteristics of real instruments. In such cases, almost all of it applies.

### 15.1. Data type/units (BTYPE/BNAME/BUNIT)

The `BTYPE`, `BNAME`, and `BUNIT` keywords should be used to describe the nature of the data, as for regular observations, see Section 15.4 " Mandatory data description keywords" and Section 18.5 "Optional data description keywords". However, simulated observations may have non-observable data, e.g.:

```
BTYPE        BNAME
ztau1        height (z-coordinate) of tau = 1 as a function of wavelength
```

### 15.2. Exposure time, binning factors

The binning factors relative to the simulation should be reported in `NBINj`, and the total binning factor in `NBIN`. For sampling, use `NSAMP` and `NSAMPj`. For instrument-mimicking sim-observations, also see [S-META].

### 15.3. Cadence

See Section 5.3, for instrument-mimicking sim-observations also Section 5.3 in [S-META].

### 15.4. Instrument/data characteristics etc.

See [S-META], and note that e.g., the `WAVExxx (WAVEMIN, WAVEMAX`, etc) keywords apply even when not mimicking a specific instrument.

#### 15.4.1. Polarimetric data

See [S-META].

Much of the contents in the corresponding Section in [S-META] is not relevant for sim-observations, such as exposure times and binning – except when emulating actual observations by an actual instrument. In such cases, all of [S-META] applies!

However, some of the keywords that may be relevant are: `CADENCE`, `WAVEMIN`, `WAVEMAX`, `WAVELNTH`, `WAVEBAND`, `POLCCONV` (for polarization data), and `DATAMIN/-MAX/-MEAN/etc`. Note that in case of variable time steps, the `CADENCE` keyword may be represented as a variable keyword plus a scalar average.

### 15.5. Quality aspects

See [S-META] for sim-observations mimicking real observations.

### 15.6. Data statistics and other properties

See Section 5.6 and Section 5.6 in [S-META].

## 16. Metadata about affiliation, origin, acquisition, etc.

See Section 6.

## 17. Grouping

*See Section 7.*

## 18. Pipeline processing applied to the data

See Section 8 in [S-META].

## 19. Integrity and administrative information

See the Section 9 in [S-META].

## 20. Reporting of events

See Section 10 i [S-META].

# Part C. Alphabetical listings of FITS keywords with section references

## 21. Alphabetical listing of all new SOLARNET keywords in this document with section references

Below is an alphabetical listing of all SOLARNET keywords in this document that are not part of the FITS standard or any widely accepted FITS convention, keywords that have been used in the past that do not have widely accepted definitions, or previously defined keywords that need to take special values in SOLARNET files:

```
BNAME          5.1., 15.1.
BTYPE          5.1., 15.1., 5.1., 15.1.
CADENCE        5.3., 5.6., 15.4.1.
ELAPSED        4., 5.6., 14.
NBIN           5.2., 15.2.
NBINj          5.2., 15.2.
NSAMP          5.2., 15.2.
NSAMPj         5.2., 15.2.
OBSRVTRY       6.
OBS_HDU        2., 12.
POINT_ID       7.
POLCCONV       15.4.1.
PROJECT        6.
SETTINGS       6.
SIM_HDU        2.
SMLATION       6., 7.
SNAPSHOT       4., 5.6., 14.
SOLARNET       2., 12., 20.
SOLNETEX       2., 12.
TARGET         6.
WAVEBAND       15.4.1.
WAVEMAX        15.4., 15.4.1.
WAVEMIN        15.4., 15.4.1.
ZTAU51         5.6.

Number of keywords: 24
```

## 22. Alphabetical listing of all keywords in this document with section references

Below is an alphabetical listing of all keywords used in this document, both the SOLARNET keywords listed in Section 21 and FITS Standard/widely accepted FITS convention keywords, the latter identified with a one-character trailing code:

       s: Keywords defined in the FITS Standard (listed in the FITS standard Appendix C)
       p: Keywords defined in Paper I-V and Thompson (2006)[1].
       o: Other keywords in common use before being specified in this document

```
BNAME          5.1., 15.1.
BTYPE          5.1., 15.1.
BUNIT      S   5.1., 15.1.
CADENCE        5.3., 5.6., 15.4.1.
CAMERA     O   6.
CAMPAIGN   O   6.
CDELTia    P   14.
CONTINUE   S   6.
CREATOR    O   6.
CRPIXja    P   14.
CRVALia    P   14.
CSYERia    P   3.1.
CTYPEia    P   14.
DATAMEAN   O   5.6.
DATAMIN    S   15.4.1.
DATE-BEG   P   4., 14.
DATEREF    P   4., 14.
ELAPSED        4., 5.6., 14.
EXTLEVEL   S   3.2.
EXTNAME    S   2., 3.2., 12.
EXTVER     S   3.2.
HGLN_OBS   P   3.1., 13.
HGLT_OBS   P   3.1., 13.
INSTRUME   S   6.
NBIN           5.2., 15.2.
NBINj          5.2., 15.2.
NSAMP          5.2., 15.2.
NSAMPj         5.2., 15.2.
OBSERVER   S   6.
OBSRVTRY       6.
OBS_HDU        2., 12.
PCi_ja     P   14.
POINT_ID       7.
POLCCONV       15.4.1.
PROJECT        6.
PSi_ma     P   3.2.
PVi_ma     P   3.2.
```

---

[1] Some of the WCS keywords defined in these papers are also defined in the FITS standard Table 22

```
RSUN_REF   P   3.1.
SETTINGS       6.
SIM_HDU        2.
SMLATION       6., 7.
SNAPSHOT       4., 5.6., 14.
SOLARNET       2., 12.
SOLNETEX       2., 12.
TARGET         6.
TFORMn     S   3.2.
TTYPEn     S   3.2.
WAVEBAND       15.4.1.
WAVELNTH   O   15.4.1.
WAVEMAX        15.4., 15.4.1.
WAVEMIN        15.4., 15.4.1.
ZTAU51         5.6.
```

**Number of keywords: 52**

## 23. Alphabetical listing of all keywords used in [S-META] with section references

Below is an alphabetical listing of all keywords used in [S-META] with corresponding section refences, both SOLARNET keywords and FITS Standard/widely accepted FITS convention keywords, the latter identified with a one-character trailing code:

> S: Keywords defined in the FITS Standard (listed in the FITS standard Appendix C)
> P: Keywords defined in Paper I-V and Thompson (2006)[1].
> O: Other keywords in common use before being specified in this document

```
ANA_NCMP       Appendix IX.
AO_LOCK        5.5., 18.6.
AO_NMODE       5.5., 18.6.
ATMOS_R0       5.5., Appendix I., Appendix I-a., Appendix I-b., 16., 18.7.
AUTHOR     S   6., 15.5.
BLANK      S   5.6.2., 15.4.
BNAME          5.1., 15.4., 18.5.
BNDCTR         5.4., 18.6.
BTYPE          5.1., 15.4.
BUNIT      S   5.1., 15.4.
BZERO      S   Appendix IV.
CADAVG         5.3., 18.3.
CADENCE        5.3., 18.3.
CADMAX         5.3., 18.3.
CADMIN         5.3., 18.3.
CADVAR         5.3., 18.3.
```

---

[1] Some of the WCS keywords defined in these papers are also defined in the FITS standard Table 22

```
CAMERA    O   6., 15.5.
CAMPAIGN  O   6., 15.5.
CAR_ROT   P   3.2.
CCURRENT      Appendix V-b., 15.5.
CDELTia   P   3.1., Appendix VI., 15.2.
CDi_ja    P   15.2.
CHECKSUM  P   9., 15.1.
CMPDESn       Appendix IX.
CMPINCn       Appendix IX.
CMPMULn       Appendix IX.
CMPNAMn       Appendix IX.
CMPSTRn       Appendix IX.
CMPTYPn       Appendix IX.
CMP_NPn       Appendix IX.
CNAMEia   P   5.1.
COMMENT   S   2., 5.5., Appendix VI-a.
COMPQUAL      5.5., 18.7.
COMP_ALG      5.5., 18.7.
CONSTEXT      Appendix IX.
CONTINUE  P   2., 2.1., 8.2., Appendix I., Appendix II., Appendix IV., 11., 15.5.,
              17., 18.4., 18.10., 18.13.
CPDISia       Appendix VI.
CPERRia       Appendix VI.
CQDISia   P   Appendix VI.
CQERRia   P   Appendix VI.
CRDERia   P   3.1., 15.2.
CREATOR   O   8.1., Appendix IV., 18.10.
CROTAia   O   3.
CRPIXja   P   3.1., 3.2., Appendix III., Appendix VI., 15.2.
CRVALia   P   3.1., Appendix VI., 15.2.
CSYERia   P   3.1., 15.2.
CTYPEia   P   3.1., 4.1., 5.1., 5.4., Appendix I., Appendix I-a., Appendix V-b.,
              Appendix IX., 15.2.
CUNITia   P   5.4., 15.2.
CWDISia       Appendix VI.
CWERRia       Appendix VI.
DATAEXT       Appendix IX.
DATAKURT  O   5.6., 18.8.
DATAMAD       5.6., 18.8.
DATAMAX   S   5.6., Appendix III., 18.8.
DATAMEAN  O   5.6., 18.8.
DATAMEDN  O   5.6., 18.8.
DATAMIN   S   5.6., Appendix III., 18.8.
DATANMAD      5.6., 18.8.
DATANPnn      5.6., 18.8.
DATANRMS      5.6., 18.8.
DATAPnn       5.6., 18.8.
DATARMS   O   5.6., 18.8.
DATASKEW  O   5.6., 18.8.
DATASUM   O   9., 15.1.
DATATAGS      6., 15.5.
DATE      S   8., 15.1.
DATE-AVG  P   4., 18.2.
DATE-BEG  P   2.2., 4., 4.1., Appendix III., 13.
DATE-END  P   4., Appendix III., 18.2.
DATEREF   P   4., 4.1., Appendix I., Appendix I-a., Appendix III., 14., 15.2.
DETECTOR  O   6., Appendix IV., 15.5.
DPja      P   Appendix VI.
```